\pdfoutput=1
\documentclass[12pt,a4paper]{article}

\usepackage{ifthen} 
\usepackage{placeins}
\usepackage{floatrow}
\usepackage{booktabs}

\newboolean{pdflatex}
\setboolean{pdflatex}{true} 

\newboolean{articletitles}
\setboolean{articletitles}{true} 

\newboolean{uprightparticles}
\setboolean{uprightparticles}{false} 

\def\paperauthors{LHCb collaboration} 
\def\paperasciititle{Evidence for modification of b quark hadronization in high-multiplicity pp collisions at sqrt(s) = 13 TeV} 
\def\papertitle{Evidence for modification of $\bquark$ quark hadronization in high-multiplicity $pp$ collisions at $\sqs = 13\tev$}
\def\paperkeywords{{High Energy Physics}, {LHCb}} 
\def\papercopyright{\the\year\ CERN for the benefit of the LHCb collaboration} 
\def\paperlicence{CC BY 4.0 licence}
\def\paperlicenceurl{https://creativecommons.org/licenses/by/4.0/}

\usepackage[top=1in, bottom=1.25in, left=1in, right=1in]{geometry}

\usepackage{microtype}
\usepackage{lineno}  
\usepackage{xspace} 
\usepackage{caption} 

\usepackage{graphicx}  
\usepackage{color}
\usepackage{colortbl}
\graphicspath{{./figs/}} 

\usepackage{amsmath} 
\usepackage{amssymb}
\usepackage{amsfonts}
\usepackage{upgreek} 

\newcommand*\patchAmsMathEnvironmentForLineno[1]{%
\expandafter\let\csname old#1\expandafter\endcsname\csname #1\endcsname
\expandafter\let\csname oldend#1\expandafter\endcsname\csname
end#1\endcsname
 \renewenvironment{#1}%
   {\linenomath\csname old#1\endcsname}%
   {\csname oldend#1\endcsname\endlinenomath}%
}
\newcommand*\patchBothAmsMathEnvironmentsForLineno[1]{%
  \patchAmsMathEnvironmentForLineno{#1}%
  \patchAmsMathEnvironmentForLineno{#1*}%
}
\AtBeginDocument{%
\patchBothAmsMathEnvironmentsForLineno{equation}%
\patchBothAmsMathEnvironmentsForLineno{align}%
\patchBothAmsMathEnvironmentsForLineno{flalign}%
\patchBothAmsMathEnvironmentsForLineno{alignat}%
\patchBothAmsMathEnvironmentsForLineno{gather}%
\patchBothAmsMathEnvironmentsForLineno{multline}%
\patchBothAmsMathEnvironmentsForLineno{eqnarray}%
}

\usepackage{hyperxmp}

\usepackage[pdftex,
            pdfauthor={\paperauthors},
            pdftitle={\paperasciititle},
            pdfkeywords={\paperkeywords},
            pdfcopyright={Copyright (C) \papercopyright},
            pdflicenseurl={\paperlicenceurl}]{hyperref}

\usepackage[colorinlistoftodos,textsize=scriptsize]{todonotes}

\usepackage[bottom,flushmargin,hang,multiple]{footmisc}

\usepackage[all]{hypcap} 

\usepackage{xspace} 
\usepackage{upgreek}

\def\lhcb   {\mbox{LHCb}\xspace}

\def\MagUp {\mbox{\em Mag\kern -0.05em Up}\xspace}

\ifthenelse{\boolean{uprightparticles}}%
{

 \def\Pmu         {\ensuremath{\upmu}\xspace}

 \def\Ppi         {\ensuremath{\uppi}\xspace}

 \def\Ppsi        {\ensuremath{\uppsi}\xspace}

 \def\PDelta      {\ensuremath{\Delta}\xspace}                 
 \def\PXi         {\ensuremath{\Xi}\xspace}                 
 \def\PLambda     {\ensuremath{\Lambda}\xspace}                 
 \def\PSigma      {\ensuremath{\Sigma}\xspace}                 
 \def\POmega      {\ensuremath{\Omega}\xspace}                 
 \def\PUpsilon    {\ensuremath{\Upsilon}\xspace}
 \let\oldPi\Pi
 \def\PPi         {\ensuremath{\oldPi}\xspace}

 \def\PB      {\ensuremath{\mathrm{B}}\xspace}                 
                  
 \def\PD      {\ensuremath{\mathrm{D}}\xspace}

 \def\PJ      {\ensuremath{\mathrm{J}}\xspace}                 
 \def\PK      {\ensuremath{\mathrm{K}}\xspace}

 \def\Pb      {\ensuremath{\mathrm{b}}\xspace}                 
                  
 \def\Pd      {\ensuremath{\mathrm{d}}\xspace}

 \def\Pi      {\ensuremath{\mathrm{i}}\xspace}

 \def\Ps      {\ensuremath{\mathrm{s}}\xspace}

 \def\thebaroffset{0.0em}
}
{

 \def\Pmu         {\ensuremath{\mu}\xspace}

 \def\Ppi         {\ensuremath{\pi}\xspace}

 \def\Ppsi        {\ensuremath{\psi}\xspace}                 
                  
 \mathchardef\PDelta="7101
 \mathchardef\PXi="7104
 \mathchardef\PLambda="7103
 \mathchardef\PSigma="7106
 \mathchardef\POmega="710A
 \mathchardef\PUpsilon="7107
 \mathchardef\PPi="7105
                  
 \def\PB      {\ensuremath{B}\xspace}                 
                  
 \def\PD      {\ensuremath{D}\xspace}

 \def\PJ      {\ensuremath{J}\xspace}                 
 \def\PK      {\ensuremath{K}\xspace}

 \def\Pb      {\ensuremath{b}\xspace}                 
                  
 \def\Pd      {\ensuremath{d}\xspace}

 \def\Pi      {\ensuremath{i}\xspace}

 \def\Ps      {\ensuremath{s}\xspace}

 \def\thebaroffset{0.18em}
}
\newcommand{\offsetoverline}[2][\thebaroffset]{\kern #1\overline{\kern -#1 #2}}%

\makeatletter
\ifcase \@ptsize \relax
  \newcommand{\miniscule}{\@setfontsize\miniscule{4}{5}}
\or
  \newcommand{\miniscule}{\@setfontsize\miniscule{5}{6}}
\or
  \newcommand{\miniscule}{\@setfontsize\miniscule{5}{6}}
\fi
\makeatother

\DeclareRobustCommand{\optbar}[1]{\shortstack{{\miniscule (\rule[.5ex]{1.25em}{.18mm})}
  \\ [-.7ex] $#1$}}

\def\mup        {{\ensuremath{\Pmu^+}}\xspace}

\def\dquark    {{\ensuremath{\Pd}}\xspace}

\def\squark    {{\ensuremath{\Ps}}\xspace}

\def\bquark    {{\ensuremath{\Pb}}\xspace}

\def\pion   {{\ensuremath{\Ppi}}\xspace}

\def\pip    {{\ensuremath{\pion^+}}\xspace}
\def\pim    {{\ensuremath{\pion^-}}\xspace}

\def\KorKbar {\kern \thebaroffset\optbar{\kern -\thebaroffset \PK}{}\xspace}

\def\D       {{\ensuremath{\PD}}\xspace}

\def\DorDbar {\kern \thebaroffset\optbar{\kern -\thebaroffset \PD}\xspace}

\def\Dp      {{\ensuremath{\D^+}}\xspace}
\def\Dm      {{\ensuremath{\D^-}}\xspace}

\def\DpDm    {\ensuremath{\Dp {\kern -0.16em \Dm}}\xspace}

\def\B       {{\ensuremath{\PB}}\xspace}

\def\BorBbar {\kern \thebaroffset\optbar{\kern -\thebaroffset \PB}\xspace}

\def\Bd      {{\ensuremath{\B^0}}\xspace}

\def\BdorBdbar {\kern \thebaroffset\optbar{\kern -\thebaroffset \Bd}\xspace}
\def\Bu      {{\ensuremath{\B^+}}\xspace}

\def\Bs      {{\ensuremath{\B^0_\squark}}\xspace}

\def\BsorBsbar {\kern \thebaroffset\optbar{\kern -\thebaroffset \Bs}\xspace}

\def\jpsi     {{\ensuremath{{\PJ\mskip -3mu/\mskip -2mu\Ppsi}}}\xspace}

\def\Upsilonres  {{\ensuremath{\PUpsilon}}\xspace}
\def\Y#1S{\ensuremath{\PUpsilon{(#1S)}}\xspace}

\def\LorLbar     {\kern \thebaroffset\optbar{\kern -\thebaroffset \PLambda}\xspace}

\def\BF         {{\ensuremath{\mathcal{B}}}\xspace}
\def\BR         {\BF}

\def\AT#1     {\ensuremath{A_{\mathrm{T}}^{#1}}\xspace}           

\def\C#1      {\ensuremath{\mathcal{C}_{#1}}\xspace}                       
\def\Cp#1     {\ensuremath{\mathcal{C}_{#1}^{'}}\xspace}                    
\def\Ceff#1   {\ensuremath{\mathcal{C}_{#1}^{\mathrm{(eff)}}}\xspace}        
\def\Cpeff#1  {\ensuremath{\mathcal{C}_{#1}^{'\mathrm{(eff)}}}\xspace}       
\def\Ope#1    {\ensuremath{\mathcal{O}_{#1}}\xspace}                       
\def\Opep#1   {\ensuremath{\mathcal{O}_{#1}^{'}}\xspace}                    

\newcommand{\aunit}[1]{\ensuremath{\text{\,#1}}}       

\newcommand{\tev}{\aunit{Te\kern -0.1em V}\xspace}
\newcommand{\gev}{\aunit{Ge\kern -0.1em V}\xspace}
\newcommand{\mev}{\aunit{Me\kern -0.1em V}\xspace}
\newcommand{\kev}{\aunit{ke\kern -0.1em V}\xspace}
\newcommand{\ev}{\aunit{e\kern -0.1em V}\xspace}
 
\newcommand{\mevc}{\ensuremath{\aunit{Me\kern -0.1em V\!/}c}\xspace}
\newcommand{\gevc}{\ensuremath{\aunit{Ge\kern -0.1em V\!/}c}\xspace}
\newcommand{\mevcc}{\ensuremath{\aunit{Me\kern -0.1em V\!/}c^2}\xspace}
\newcommand{\gevcc}{\ensuremath{\aunit{Ge\kern -0.1em V\!/}c^2}\xspace}

\def\gsim{{~\raise.15em\hbox{$>$}\kern-.85em
          \lower.35em\hbox{$\sim$}~}\xspace}
\def\lsim{{~\raise.15em\hbox{$<$}\kern-.85em
          \lower.35em\hbox{$\sim$}~}\xspace}

\def\sPlot{\mbox{\em sPlot}\xspace}

\def\sqs   {\ensuremath{\protect\sqrt{s}}\xspace}

\def\pt         {\ensuremath{p_{\mathrm{T}}}\xspace}

\def\evtgen     {\mbox{\textsc{EvtGen}}\xspace}

\def\geant      {\mbox{\textsc{Geant4}}\xspace}

\def\pythia     {\mbox{\textsc{Pythia}}\xspace}

\def\tell1  {TELL1\xspace}
\def\ukl1   {UKL1\xspace}

\newcommand{\lhcborcid}[1]{\href{https://orcid.org/#1}{\hspace*{0.1em}\raisebox{-0.45ex}{\includegraphics[width=1em]{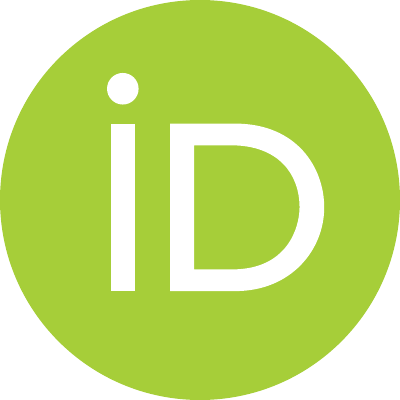}}}}

\usepackage{cite} 
\usepackage{mciteplus}

\def\Nvelo {\ensuremath{\mathrm{N}_{\mathrm{tracks}}^{\mathrm{VELO}}}\xspace}
\def\Nback {\ensuremath{\mathrm{N}_{\mathrm{tracks}}^{\mathrm{back}}}\xspace}

\usepackage{longtable} 

\begin{document}

\renewcommand{\thefootnote}{\fnsymbol{footnote}}
\setcounter{footnote}{1}


\begin{titlepage}
\pagenumbering{roman}

\vspace*{-1.5cm}
\centerline{\large EUROPEAN ORGANIZATION FOR NUCLEAR RESEARCH (CERN)}
\vspace*{1.5cm}
\noindent
\begin{tabular*}{\linewidth}{lc@{\extracolsep{\fill}}r@{\extracolsep{0pt}}}
\ifthenelse{\boolean{pdflatex}}
{\vspace*{-1.5cm}\mbox{\!\!\!\includegraphics[width=.14\textwidth]{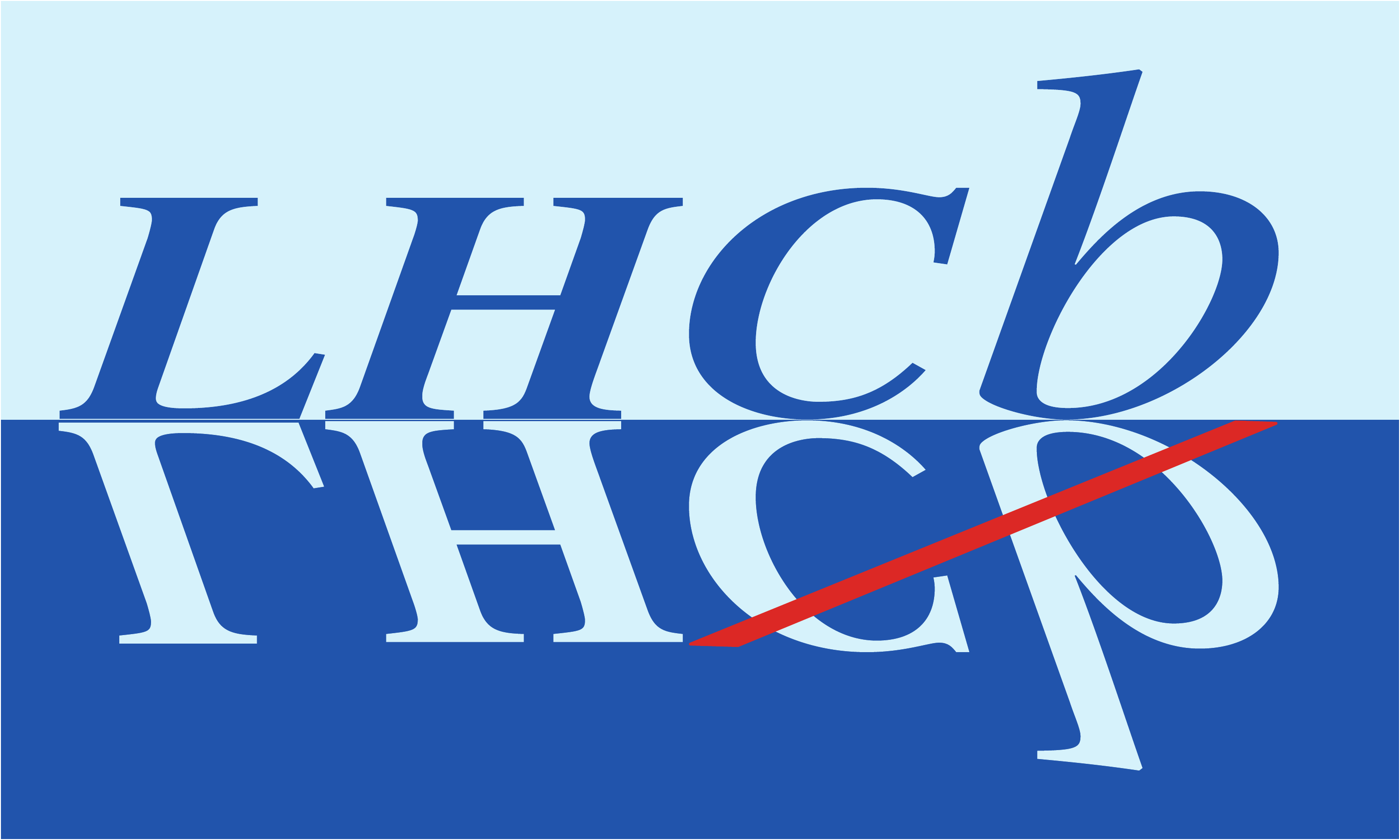}} & &}%
{\vspace*{-1.2cm}\mbox{\!\!\!\includegraphics[width=.12\textwidth]{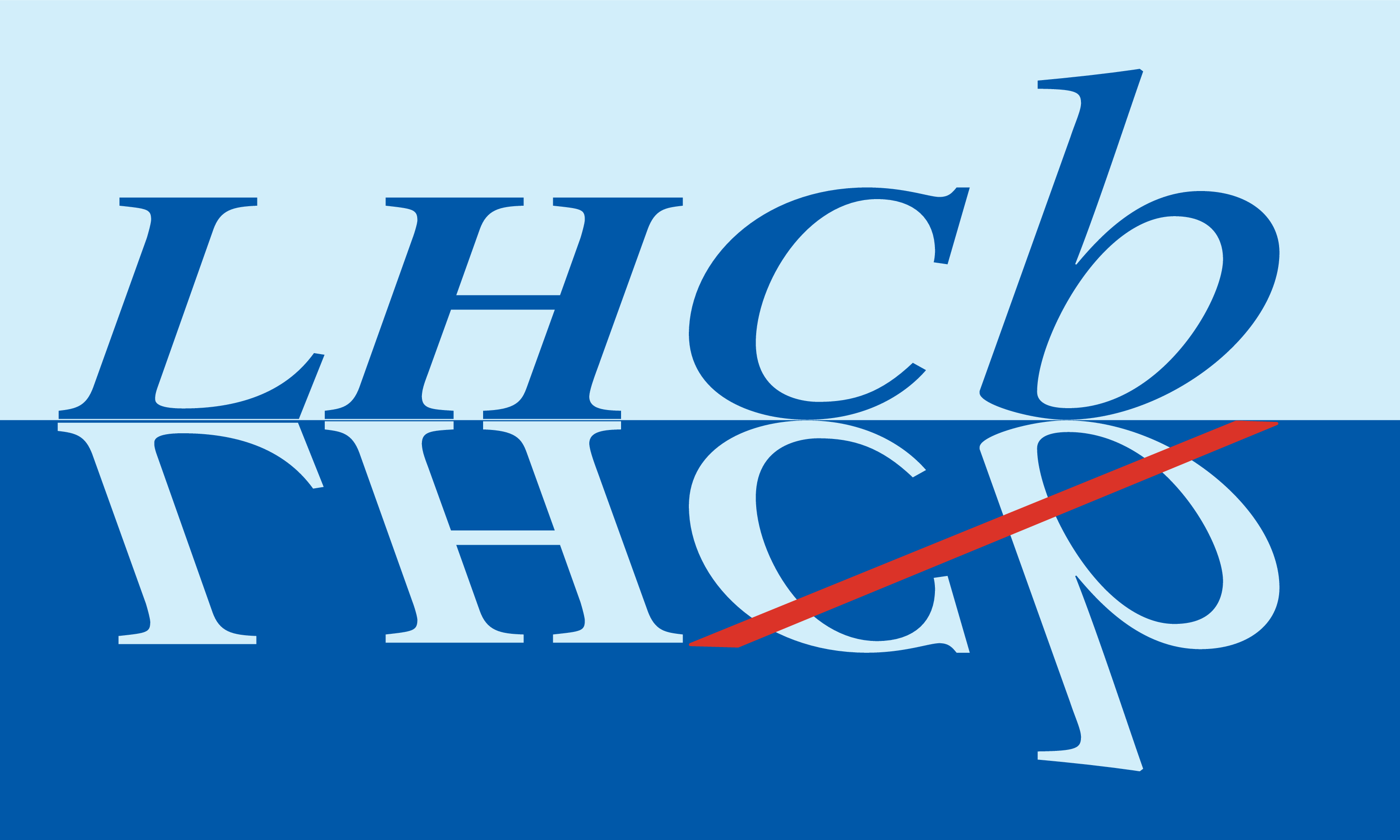}} & &}%
\\
 & & CERN-EP-2022-062 \\  
 & & LHCb-PAPER-2022-001 \\  
 & & \today \\ 
 & & \\
\end{tabular*}

\vspace*{4.0cm}

{\normalfont\bfseries\boldmath\huge
\begin{center}
  \papertitle 
\end{center}
}

\vspace*{2.0cm}
\begin{center}
\paperauthors
\footnote{Authors are listed at the end of this Letter.}
\end{center}

\vspace{\fill}

\begin{abstract}
  \noindent
The production rate of $\Bs$ mesons relative to $\Bd$ mesons is measured by the LHCb experiment in $pp$ collisions at a center-of-mass energy $\sqrt{s} = 13\tev$ over the forward rapidity interval $2<y<4.5$ as a function of the charged particle multiplicity measured in the event.  Evidence at the 3.4$\sigma$ level is found for an increase of the ratio of $\Bs$ to $\Bd$ cross-sections with multiplicity at transverse momenta below 6 GeV/$c$, with no significant multiplicity dependence at higher transverse momentum.  Comparison with data from $e^{+}e^{-}$ collisions implies that the density of the hadronic medium may affect the production rates of $B$ mesons. This is qualitatively consistent with the emergence of quark coalescence as an additional hadronization mechanism in high-multiplicity collisions.
  
\end{abstract}

\vspace*{2.0cm}

\begin{center}
Published in Phys.~Rev.~Lett. 131 (2023) 061901
\end{center}

\vspace{\fill}

{\footnotesize 
\centerline{\copyright~\papercopyright. \href{\paperlicenceurl}{\paperlicence}.}}
\vspace*{2mm}

\end{titlepage}


\newpage
\setcounter{page}{2}
\mbox{~}
%
%
%
%


\renewcommand{\thefootnote}{\arabic{footnote}}
\setcounter{footnote}{0}

\cleardoublepage


\pagestyle{plain} 
\setcounter{page}{1}
\pagenumbering{arabic}


Measurements of $B$ mesons at colliders offer unique probes of the hadronization process by which single quarks evolve into color-neutral hadrons.  In contrast to light quarks, the large mass of $\bquark$ quarks suppresses their production via non-perturbative processes. In addition there is no $\bquark$ content in the valence quark distribution of the incoming beam particles \cite{Norrbin:2000zc}.  Therefore, production of $b\bar{b}$ pairs at hadron colliders is dominated by hard parton-parton interactions in the initial stages of the collisions, and is well described by perturbative QCD calculations \cite{LHCb-PAPER-2016-031,LHCb-PAPER-2017-037,ALICE:2021mgk}.  

The fraction of $\bquark$ quarks that pair with an $\squark$ quark to form $\Bs$ mesons, $f_{s}$, and the fraction that pair with a light $\dquark$ quark to form $\Bd$ mesons, $f_{d}$, are determined through the hadronization process.  One mechanism for hadronization is fragmentation, where showers of partons produced by outgoing quarks form into hadrons \cite{Andersson:1983ia,Webber:1983if}. Measurements of $B$ hadron production in $e^{+}e^{-}$ collisions at the $\Upsilonres(5S)$ \cite{CLEO:2005pyn,Belle:2006jvm,CLEO:2006dkk} and $Z^{0}$ \cite{OPAL:1992zsd, ALEPH:1995npv,L3:2000gcz, DELPHI:2003pao} resonances give consistent values for the ratio $f_{s}/f_{d}$, which is often interpreted as evidence for the universality of $\bquark$ quark fragmentation assumed by QCD factorization theorems \cite{Collins:1989gx}.  However, measurements at hadron colliders have shown that the ratio $f_{s}/f_{d}$ has a dependence on the collision center-of-mass energy and the $B$ meson transverse momentum $\pt$ \cite{CDF:1996ivk,LHCb-PAPER-2012-037, ATLAS:2015esn, LHCb-PAPER-2019-020,LHCb-PAPER-2020-046}. The fraction of $\bquark$ quarks which hadronize into baryons also varies with $\pt$ \cite{LHCb-PAPER-2011-018,LHCb-PAPER-2018-050}.  Additionally, recent measurements have shown that charm quark hadronization differs between $e^{+}e^{-}$ and $pp$ collisions \cite{ALICE:2021dhb}. The reason for these variations is not immediately clear, and may be explained by hadronization mechanisms other than fragmentation \cite{Berezhnoy:2013qln}.

An alternative hadronization process, quark coalescence, can occur when a quark produced in the collision combines with another quark to form a color singlet hadron.  Models incorporating coalescence are successful at reproducing a range of measurements from fixed-target experiments and colliders \cite{Vogt:1994zf,Braaten:2002yt,Rapp:2003wn,He:2019vgs,Minissale:2020bif}. Coalescence calculations generally require multiple quark wavefunctions to overlap in position and velocity, so the fraction of hadrons produced by this mechanism is expected to increase with the number of quarks produced in the collision.  The effect is expected to be most prominent at relatively low $\pt$, which is the range where the bulk of the particles created in the collision are found.  Coalescence can also lead to enhanced production of baryons at low $\pt$, and is especially important in high-energy heavy ion collisions where a large volume of deconfined quark-gluon plasma (QGP) is formed \cite{Fries:2003vb,Greco:2003xt,Molnar:2003ff}. Data from the CMS collaboration has shown that $\Bs$ production may be enhanced relative to $\Bu$ production in PbPb collisions when compared to $pp$ collisions \cite{CMS:2018eso,CMS:2021mzx}. However, significant uncertainties and the relatively high $\pt$ range covered by that data preclude drawing firm conclusions on the influence of coalescence on $B$ hadronization.

Recent measurements in $pp$ collisions have shown some behaviors similar to those associated with the formation of QGP in collisions of heavy nuclei \cite{CMS:2010ifv,ATLAS:2015hzw,CMS:2016fnw}.  Among these effects is an enhanced yield of light-quark baryons and mesons with strangeness in collisions where a relatively large number of charged particles are produced \cite{ALICE:2017jyt}, which was originally proposed as a QGP signature \cite{Koch:1986ud}.  If hadronization via coalescence emerges as a mechanism for forming final state $B$ hadrons, then the production rates of $\Bs$ hadrons could increase relative to the production of $\Bd$ hadrons as particle multiplicity increases. 

This Letter describes LHCb measurements of the ratio of $\Bs$ to $\Bd$ cross sections, $\sigma_{\Bs}/\sigma_{\Bd}$,  as a function of charged particle multiplicity and $\pt$.  Both the $\Bs$ and $\Bd$ candidates are reconstructed through their decays to the $\jpsi \pip \pim$ final state, where the $\jpsi$ decays into a $\mu^{+}\mu^{-}$ pair.  This decay mode provides similar yields for both $\Bs$ and $\Bd$ mesons.
Here multiplicity is represented by the number of charged tracks reconstructed in a silicon strip detector that surrounds the $pp$ interaction region, the LHCb VELO detector \cite{LHCb-TDR-005,LHCb-DP-2014-001}.  These measurements use a sample of $pp$ collisions collected at a center-of-mass energy $\sqrt{{s}} = 13\tev$, corresponding to an integrated luminosity of 5.4~fb$^{-1}$.

The LHCb detector is a single-arm forward spectrometer covering the pseudorapidity range $2 < \eta < 5$, described in detail in Refs.~\cite{LHCb-DP-2008-001,LHCb-DP-2014-002}.  Events considered in this analysis are required to satisfy a series of triggers designed to select the decay $\jpsi\rightarrow\mup\mu^{-}$ and have one reconstructed $pp$ interaction point (primary vertex).  Each muon candidate is required to penetrate the hadron absorber layers in the LHCb muon system and have $\pt> 500 \mevc$.  Candidate $\jpsi$ mesons are formed from pairs of oppositely charged muon candidates that have an invariant mass near the known $\jpsi$ mass and originate from a vertex that is displaced from the primary vertex.  Charged pion candidates are identified by the response of the LHCb ring-imaging Cherenkov detectors, and are required to have total momentum $p>3\gevc$ and transverse momentum $\pt>750\mevc$.  Candidate $\mup\mu^{-}\pip\pim$ combinations that form a good quality common vertex are retained, and the tracks are refit with kinematic constraints that fix the $\mup\mu^{-}$ invariant mass to the known $\jpsi$ mass, and require all four tracks to have the same origin point \cite{Hulsbergen:2005pu}.  

Simulation is required to model the effects of the detector acceptance and the selection requirements.  In the simulation, $pp$ collisions are generated using \pythia~\cite{Sjostrand:2007gs,*Sjostrand:2006za} with a specific \lhcb configuration~\cite{LHCb-PROC-2010-056}. Decays of unstable particles are described by \evtgen~\cite{Lange:2001uf}.  The interaction of the generated particles with the detector, and its response, are implemented using the \geant toolkit~\cite{Allison:2006ve, *Agostinelli:2002hh} as described in Ref.~\cite{LHCb-PROC-2011-006}. The $\pt$ distributions of the simulated $\Bs$ and $\Bd$ mesons, the invariant mass distributions of $\pip \pim$ pairs from their decays, and simulated event multiplicity distributions are weighted to match background-subtracted distributions that are extracted from the data using the \sPlot method \cite{splot}. 

\begin{figure}[!ht]
  \centering
  \includegraphics[width=0.7\linewidth]{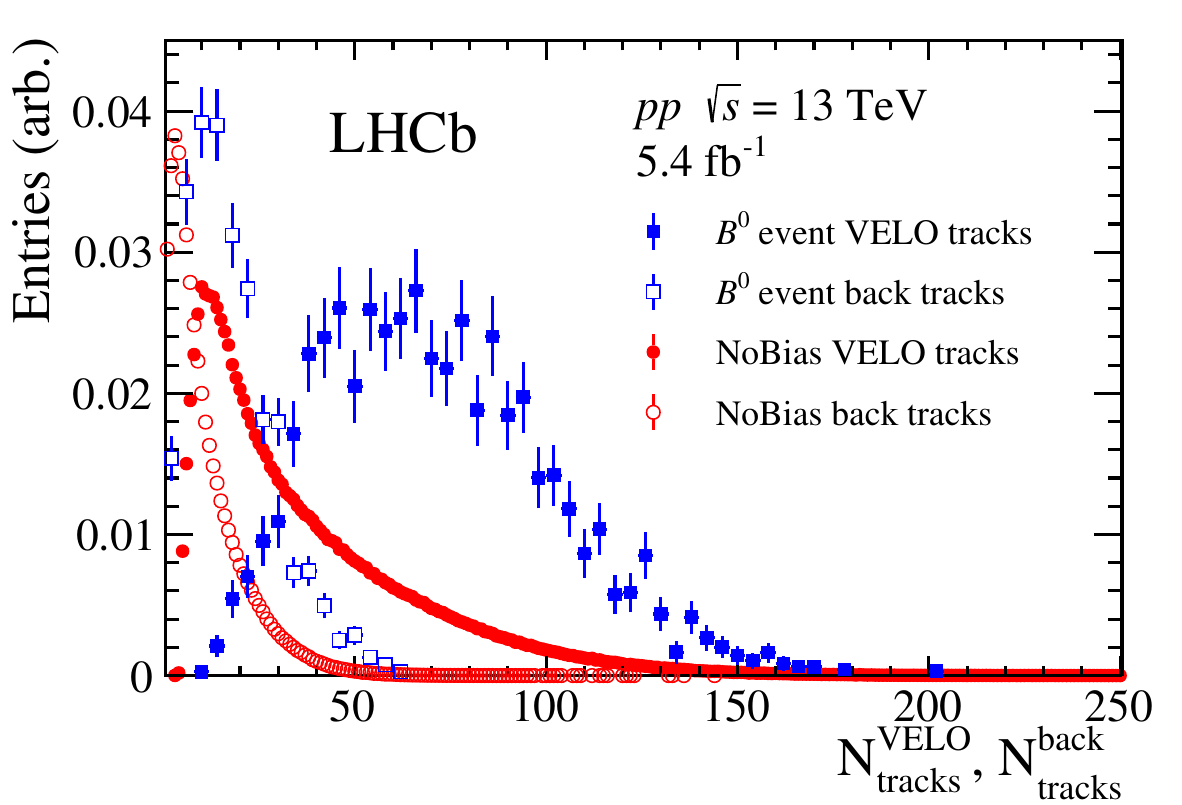}
  \caption{Distribution of the number of VELO tracks and backward tracks for NoBias events (red) and $\Bd$ signal events (blue), each with only one primary vertex. The vertical scale is arbitrary.}
  \label{fig:Fig1}
\end{figure}

The multiplicity metrics used in this analysis are the total number of charged tracks reconstructed in the VELO detector, $\Nvelo$, and the subset of VELO tracks that point in the backward direction, away from the LHCb spectrometer, $\Nback$.  The backward tracks cover a pseudorapidity interval of approximately $-3.5<\eta<-1.5$, providing a multiplicity estimate that is measured in a different region than the signal.  The VELO detector and its performance are described in detail in Refs. \cite{LHCb-TDR-005,LHCb-DP-2014-001}.  Figure \ref{fig:Fig1} shows the distributions of $\Nvelo$ and $\Nback$ for both NoBias events and $\Bd$ signal events with one reconstructed primary vertex, which requires at least five reconstructed tracks. NoBias events are selected based on the Large Hadron Collider beam clock, which indicates that a bunch crossing has occurred, without any other trigger requirements.  The distributions for $\Bd$ signal events are extracted from the data, and background is removed using the \sPlot method \cite{splot}.  The results are quoted in terms of normalized multiplicity, defined as the number of tracks at the center of a given multiplicity interval divided by the mean number of tracks in NoBias events, which are $\langle \Nvelo \rangle$$_{\mathrm{NoBias}}$ = 37.7 and $\langle \Nback \rangle$$_{\mathrm{NoBias}}$ = 11.1, with negligibly small uncertainties.  For comparison, the mean number of $\Nvelo$ and $\Nback$ are $71.1\pm0.1$ and $17.4\pm0.3$ for $\Bd$ signal events, respectively, where the uncertainty is due to the statistical uncertainty on the track distributions.  In some respects, the low- and high-multiplicity data samples approach the hadronic environments achieved in $e^{+}e^{-}$ collisions and heavy-ion collisions, respectively.

\begin{figure}[b]
  \centering
  \includegraphics[width=1.0\linewidth]{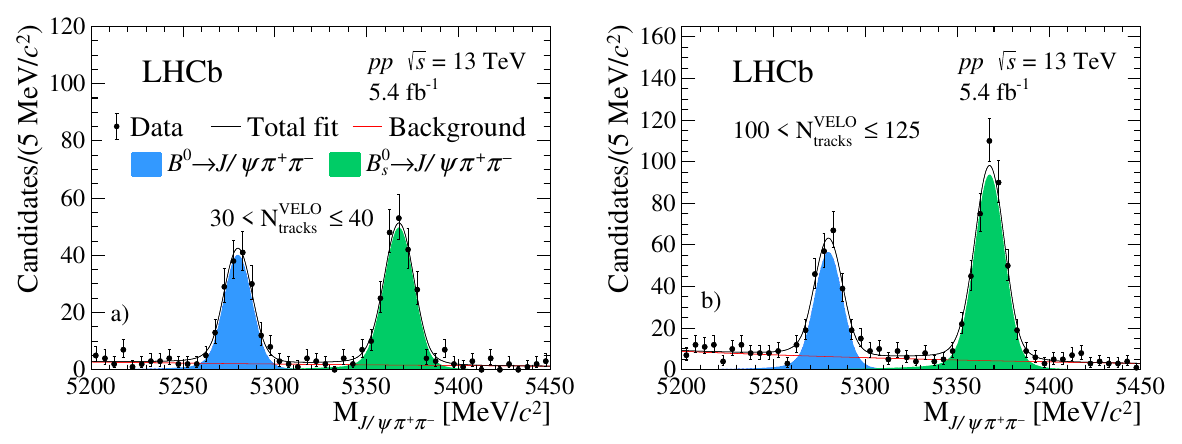}
  \caption{Measured $\jpsi \pip \pim$ invariant mass distributions and fit projections in the multiplicity ranges a) $30<\Nvelo\leq40$ and b) $100<\Nvelo\leq125$.}
  \label{fig:Bs_fit_example}
\end{figure}

The sample containing signal events is divided into intervals of multiplicity, and in each interval a likelihood fit is performed on the $\jpsi \pip \pim$ invariant mass spectrum to determine the ratio of $\Bs$ to $\Bd$ yields.  Examples of the $\jpsi \pip \pim$ mass distribution in low- and high-multiplicity intervals are shown in Fig. \ref{fig:Bs_fit_example}.  An increase of the $\Bs$ yield relative to the $\Bd$ yield in the high-multiplicity interval is apparent.  The $\Bs$ and $\Bd$ peaks are each represented in the fit by a sum of two Crystal Ball functions, which have tail parameters constrained to values determined by simulation.  The background contribution is represented by an exponential function, which is found to provide a good description of the purely combinatorial $\jpsi \pi^{\pm} \pi^{\pm}$ mass spectrum with like-sign dipions.  All multiplicity intervals are fit simultaneously, where the signal shapes are constrained to be the same in each interval, but their normalization and the background parameters are allowed to vary.  The $\Bs$ and $\Bd$ line shapes are nearly identical, and variations of the fit functions have a negligible effect on the extracted ratio of $\Bs$ to $\Bd$ yields. 

 The ratio of cross-sections $\sigma_{\Bs}/\sigma_{\Bd}$ is found by calculating
\begin{equation}
\label{eqn:rat}
\begin{aligned}
\frac{\sigma_{\Bs}}{\sigma_{\Bd}} = \frac{N_{\Bs}}{N_{\Bd}} \times
\frac{\BR_{\Bd}}{\BR_{\Bs}} \times
\frac{\varepsilon_{\Bd}^{\mathrm{acc}}}{\varepsilon_{\Bs}^{\mathrm{acc}}} \times
\frac{\varepsilon_{\Bd}^{\mathrm{trig}}}{\varepsilon_{\Bs}^{\mathrm{trig}}} \times
\frac{\varepsilon_{\Bd}^{\mathrm{PID}}}{\varepsilon_{\Bs}^{\mathrm{PID}}} \times
\frac{\varepsilon_{\Bd}^{\mathrm{reco}}}{\varepsilon_{\Bs}^{\mathrm{reco}}},\\
\end{aligned}
\end{equation}
where $N_{\Bs}/N_{\Bd}$ is the ratio of $\Bs$ to $\Bd$ signal yields returned by the fit, $\BR_{\Bd}/\BR_{\Bs}$ is the ratio of $\Bd$ to $\Bs$ branching fractions to $\jpsi \pip \pim$ \cite{ParticleDataGroup:2020ssz,LHCb-PAPER-2020-046}, and $\varepsilon_{\Bd}^{\mathrm{acc}}/\varepsilon_{\Bs}^{\mathrm{acc}},
\varepsilon_{\Bd}^{\mathrm{trig}}/\varepsilon_{\Bs}^{\mathrm{trig}},
\varepsilon_{\Bd}^{\mathrm{PID}}/\varepsilon_{\Bs}^{\mathrm{PID}}$, and  $\varepsilon_{\Bd}^{\mathrm{reco}}/\varepsilon_{\Bs}^{\mathrm{reco}}$ are ratios of the LHCb acceptance and the trigger, particle identification, and reconstruction efficiencies for $\Bd$ to $\Bs$ mesons, respectively.  Due to the similarities of the $\Bs$ and $\Bd$ decays, many systematic uncertainties partially cancel in this ratio of cross sections.  The ratio of the LHCb acceptance for the decays $\varepsilon_{\Bd}^{\mathrm{acc}}/\varepsilon_{\Bs}^{\mathrm{acc}}$ is found, using simulation, to be consistent with unity, with an uncertainty of $\sim1\%$ due to the uncertainty on the weights applied to the simulation in order to match the data.  The ratio of trigger efficiencies $\varepsilon_{\Bd}^{\mathrm{trig}}/\varepsilon_{\Bs}^{\mathrm{trig}}$ is determined from data to be consistent with unity, with an uncertainty of $\sim1\%$, using techniques described in Ref. \cite{LHCb-PUB-2014-039}, where the uncertainty comes from statistical uncertainties on the data sample. The ratio of particle identification efficiencies $\varepsilon_{\Bd}^{\mathrm{PID}}/\varepsilon_{\Bs}^{\mathrm{PID}}$ is found using calibrated samples of identified muons and pions obtained from the data, and is consistent with unity with an uncertainty of $\sim1\%$ due to the finite size of the calibration sample.  The only term with a significant difference from unity is the ratio of reconstruction efficiencies, which is found to be \mbox{ $\varepsilon_{\Bd}^{\mathrm{reco}}/\varepsilon_{\Bs}^{\mathrm{reco}}$ = $0.86\pm0.04$} for the $\pt$-integrated sample.  This is due to the difference in the dipion mass distributions produced in the $\Bs$ and $\Bd$ decays: the $\Bs$ decay is dominated by contributions from intermediate $f_{0}(980)$ and $f_{0}(1500)$ states \cite{LHCb-PAPER-2013-069}, which are reconstructed with higher efficiency than the lower-mass $\rho^{0}(770)$ intermediate state that is significant in $\Bd$ decays \cite{LHCb-PAPER-2014-012}.  The uncertainty on this correction is due to the statistical uncertainty on the weights extracted from the data that are applied to the simulation in order to match the measured $B$ meson $\pt$ and dipion mass distributions.

The ratio of cross-sections for the multiplicity-integrated samples is found to be $\sigma_{\Bs}/\sigma_{\Bd} = 0.30\pm 0.01\pm 0.03$, where the first uncertainty is statistical and the second is systematic.  This measurement agrees with previous LHCb measurements of $f_{s}/f_{d}$ using different decay channels \cite{LHCb-PAPER-2020-046} within $1.5$ standard deviations.

\begin{figure}[b]
  \centering
  \includegraphics[width=0.8\linewidth]{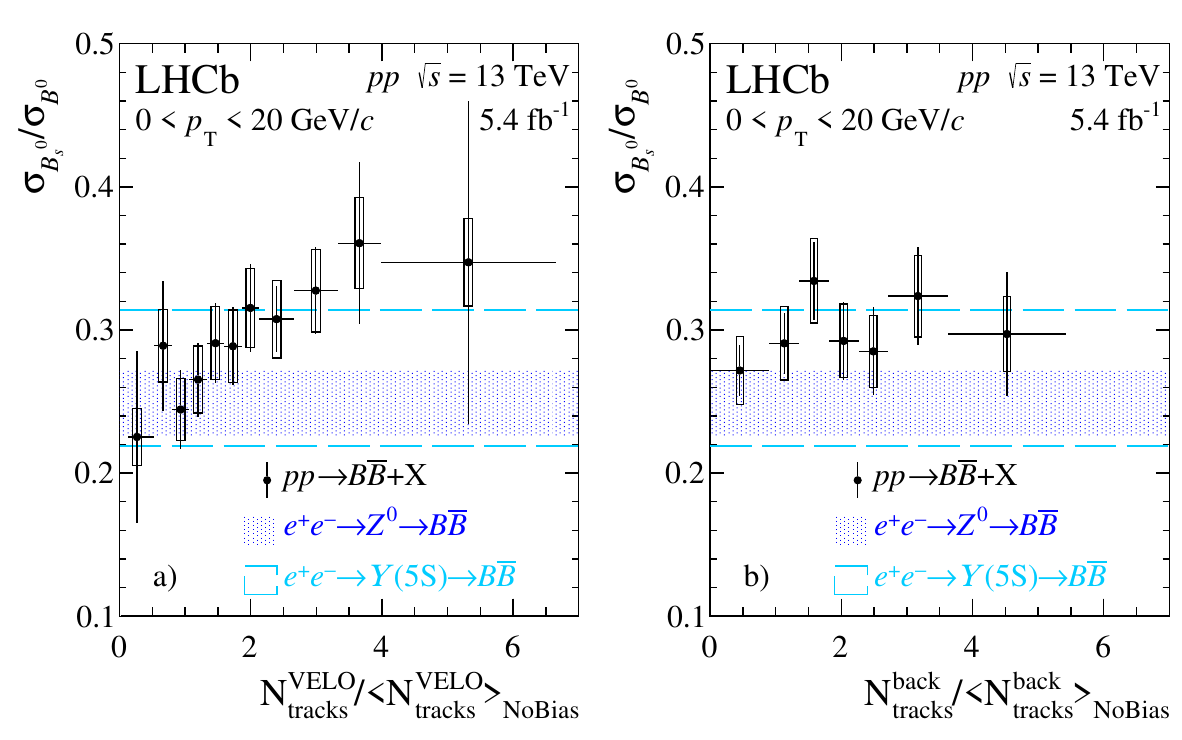}
  \caption{Ratio of cross sections $\sigma_{\Bs}/\sigma_{\Bd}$ versus the normalized multiplicity of a) all VELO tracks, and b) backward VELO tracks.  The vertical error bars (boxes) represent point-to-point uncorrelated (fully correlated) uncertainties.  The horizontal bands show the values measured in $e^{+}e^{-}$ collisions.}
  \label{fig:Fig3}
\end{figure}

The multiplicity dependence of $\sigma_{\Bs}/\sigma_{\Bd}$ is shown in Fig. \ref{fig:Fig3}, for two different multiplicity metrics.  The vertical error bars (boxes) represent point-to-point uncorrelated (fully correlated) uncertainties, while the horizontal error bars represent the bin width.  Numerical values are given in the supplemental material~\cite{ref:SuppMat}.  In the left panel, the ratio shows an increasing trend with the total VELO multiplicity, where multiplicity is normalized to the mean value found in NoBias collisions.  Also shown are the $\sigma_{\Bs}/\sigma_{\Bd}$ values measured in $e^{+}e^{-}$ collisions at the $\Upsilonres(5S)$ and $Z^{0}$ resonances \cite{HFLAV:2019otj}, which are in good agreement with the data at low multiplicity.  The right panel shows the same ratio versus the normalized $\Nback$.  No significant dependence is observed on the multiplicity measured in the backward region. The dependence on total multiplicity, compared to the lack of dependence on multiplicity measured at backward rapidity, could indicate that the mechanism responsible for the increase in the $\sigma_{\Bs}/\sigma_{\Bd}$ ratio is related to the local particle density in a similar rapidity interval as the $B$ mesons themselves.

The multiplicity dependence of $\sigma_{\Bs}/\sigma_{\Bd}$ is shown in three different intervals of $B$ meson $\pt$ in Fig. \ref{fig:Fig4}.  Numerical values are given in the supplemental material~\cite{ref:SuppMat}.  The lowest $\pt$ interval, $0<\pt<6$ GeV/$c$, encompasses $B$ mesons with $\pt$ approximately equal to or less than their mass.  In this $\pt$ interval, at low multiplicity the $\sigma_{\Bs}/\sigma_{\Bd}$ ratio is consistent with values measured in $e^{+}e^{-}$ collisions, and increases with multiplicity. A line fit to these data returns a slope of $0.075\pm0.022$ $(\Nvelo /\langle \Nvelo \rangle_{\mathrm{NoBias}})^{-1}$, which differs from zero by 3.4 standard deviations and thereby provides evidence for an increase of the the $\sigma_{\Bs}/\sigma_{\Bd}$ ratio. This fit considers only the point-to-point uncorrelated uncertainties; since all data points must move simultaneously within the correlated uncertainties, they have no effect on the extracted slope.

For comparison, the ratio of cross sections calculated by the PYTHIA event generator is also shown.  Events are generated with and without color reconnection (CR), a process which allows partons produced by from different interactions in the collision to be connected by color lines \cite{Sjostrand:1987su,Sjostrand:2013cya}.  Color reconnection was introduced to model charged particle production in hadron collisions, and is thought to be especially important in high-multiplicity collisions where multiple-parton interactions are significant.  Both sets of PYTHIA calculations show a rise with multiplicity,  which is more pronounced when CR is included.  The PYTHIA models agree with the data at relatively low multiplicity, but both scenarios give values that are lower than the central values of the data at high multiplicity.

The measurements in higher $\pt$ intervals, $6<\pt<12$ GeV/$c$ and $12<\pt<20$ GeV/$c$, display no significant dependence on multiplicity and are consistent with data from $e^{+}e^{-}$ collisions.  This behavior is expected in a scenario where low-$\pt$ $\bquark$ quarks with relatively low velocity recombine with $s$ quarks produced in high-multiplicity collisions, while the wavefunctions of higher $\pt$ $\bquark$ quarks have less overlap with the low-$\pt$ bulk of the quarks produced in the collision.  These high-$\pt$ $\bquark$ quarks would thereby dominantly hadronize via fragmentation in vacuum, as in $e^{+}e^{-}$ collisions, rather than via coalescence.  Again, both sets of PYTHIA calculations show a rising trend, which is more pronounced when color reconnection is included.  However, here the uncertainties on the data prevent discrimination between the two scenarios.  Lines fit to the $6<\pt<12$ GeV/$c$ and $12<\pt<20$ GeV/$c$ data have slopes that are consistent with zero within 0.2 and 1.3 standard deviations, respectively.

\begin{figure}[t]
  \centering
  \includegraphics[width=1.0\linewidth]{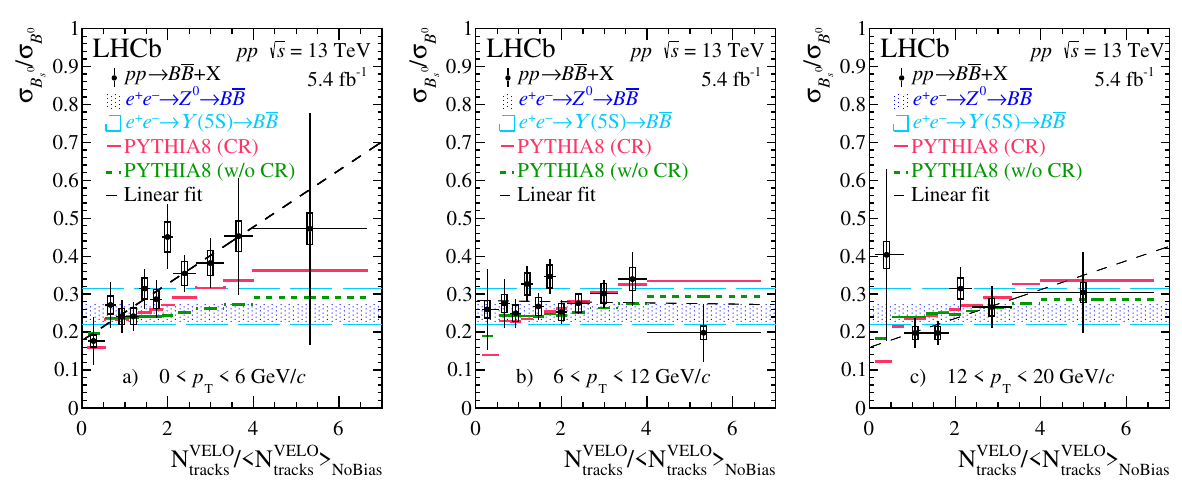}
  \caption{Ratio of cross sections $\sigma_{\Bs}/\sigma_{\Bd}$ versus normalized multiplicity in the transverse momentum ranges a) $0<\pt<6$ GeV/$c$, b) $6<\pt<12$ GeV/$c$, and c) $12<\pt<20$ GeV/$c$.  The vertical error bars (boxes) represent point-to-point uncorrelated (fully correlated) uncertainties.  The horizontal bands show the values measured in $e^{+}e^{-}$ collisions and the horizontal lines show PYTHIA calculations with and without (w/o) color reconnection.}
  \label{fig:Fig4}
\end{figure}

In summary, LHCb measurements in $pp$ collisions at $\sqs$ = 13\tev show evidence that the production of $\Bs$ mesons is enhanced relative to $\Bd$ mesons in collisions with high charged-particle multiplicity, indicating that strangeness enhancement is present in $B$ hadron production.  In collisions with relatively low charged-particle multiplicity, and for $B$ mesons with $\pt>6$ GeV/$c$, the rate of $\Bs$ production relative to $\Bd$ production is consistent with what is measured in $e^+e^-$ collisions.  These measurements are qualitatively consistent with expectations based on the emergence of quark coalescence as an additional hadronization mechanism, rather than fragmentation alone.  These results could indicate that interactions of the $\bquark$ quarks with the local hadronic environment influence the hadronization process, thereby breaking factorization of $\bquark$ quark hadronization between $e^{+}e^{-}$ and hadron collisions.

\section*{Acknowledgements}
%
%
\noindent We express our gratitude to our colleagues in the CERN
accelerator departments for the excellent performance of the LHC. We
thank the technical and administrative staff at the LHCb
institutes.
We acknowledge support from CERN and from the national agencies:
CAPES, CNPq, FAPERJ and FINEP (Brazil); 
MOST and NSFC (China); 
CNRS/IN2P3 (France); 
BMBF, DFG and MPG (Germany); 
INFN (Italy); 
NWO (Netherlands); 
MNiSW and NCN (Poland); 
MEN/IFA (Romania); 
MICINN (Spain); 
SNSF and SER (Switzerland); 
NASU (Ukraine); 
STFC (United Kingdom); 
DOE NP and NSF (USA).
We acknowledge the computing resources that are provided by CERN, IN2P3
(France), KIT and DESY (Germany), INFN (Italy), SURF (Netherlands),
PIC (Spain), GridPP (United Kingdom), 
CSCS (Switzerland), IFIN-HH (Romania), CBPF (Brazil),
Polish WLCG  (Poland) and NERSC (USA).
We are indebted to the communities behind the multiple open-source
software packages on which we depend.
Individual groups or members have received support from
ARC and ARDC (Australia);
Minciencias (Colombia);
AvH Foundation (Germany);
EPLANET, Marie Sk\l{}odowska-Curie Actions and ERC (European Union);
A*MIDEX, ANR, IPhU and Labex P2IO, and R\'{e}gion Auvergne-Rh\^{o}ne-Alpes (France);
Key Research Program of Frontier Sciences of CAS, CAS PIFI, CAS CCEPP, 
Fundamental Research Funds for the Central Universities, 
and Sci. \& Tech. Program of Guangzhou (China);
GVA, XuntaGal, GENCAT and Prog.~Atracci\'on Talento, CM (Spain);
SRC (Sweden);
the Leverhulme Trust, the Royal Society
 and UKRI (United Kingdom).

\addcontentsline{toc}{section}{References}
\bibliographystyle{LHCb}
\bibliography{main,standard,LHCb-PAPER,LHCb-CONF,LHCb-DP,LHCb-TDR}

\ifx\mcitethebibliography\mciteundefinedmacro
\PackageError{LHCb.bst}{mciteplus.sty has not been loaded}
{This bibstyle requires the use of the mciteplus package.}\fi
\providecommand{\href}[2]{#2}
\begin{mcitethebibliography}{10}
\mciteSetBstSublistMode{n}
\mciteSetBstMaxWidthForm{subitem}{\alph{mcitesubitemcount})}
\mciteSetBstSublistLabelBeginEnd{\mcitemaxwidthsubitemform\space}
{\relax}{\relax}

\bibitem{Norrbin:2000zc}
E.~Norrbin and T.~Sjostrand,
  \ifthenelse{\boolean{articletitles}}{\emph{{Production and hadronization of
  heavy quarks}}, }{}\href{https://doi.org/10.1007/s100520000460}{Eur.\ Phys.\
  J.\  \textbf{C17} (2000) 137},
  \href{http://arxiv.org/abs/hep-ph/0005110}{{\normalfont\ttfamily
  arXiv:hep-ph/0005110}}\relax
\mciteBstWouldAddEndPuncttrue
\mciteSetBstMidEndSepPunct{\mcitedefaultmidpunct}
{\mcitedefaultendpunct}{\mcitedefaultseppunct}\relax
\EndOfBibitem
\bibitem{LHCb-PAPER-2016-031}
LHCb collaboration, R.~Aaij {\em et~al.},
  \ifthenelse{\boolean{articletitles}}{\emph{{Measurement of the \bquark-quark
  production cross-section in 7 and 13$\tev$ $\proton\proton$ collisions}},
  }{}\href{https://doi.org/10.1103/PhysRevLett.118.052002}{Phys.\ Rev.\ Lett.\
  \textbf{118} (2017) 052002}, Erratum
  \href{https://doi.org/10.1103/PhysRevLett.119.169901}{ibid.\   \textbf{119}
  (2017) 169901}, \href{http://arxiv.org/abs/1612.05140}{{\normalfont\ttfamily
  arXiv:1612.05140}}\relax
\mciteBstWouldAddEndPuncttrue
\mciteSetBstMidEndSepPunct{\mcitedefaultmidpunct}
{\mcitedefaultendpunct}{\mcitedefaultseppunct}\relax
\EndOfBibitem
\bibitem{LHCb-PAPER-2017-037}
LHCb collaboration, R.~Aaij {\em et~al.},
  \ifthenelse{\boolean{articletitles}}{\emph{{Measurement of the \Bpm
  production cross-section in \proton\proton collisions at $\sqs = $7 and
  13\tev}}, }{}\href{https://doi.org/10.1007/JHEP12(2017)026}{JHEP \textbf{12}
  (2017) 026}, \href{http://arxiv.org/abs/1710.04921}{{\normalfont\ttfamily
  arXiv:1710.04921}}\relax
\mciteBstWouldAddEndPuncttrue
\mciteSetBstMidEndSepPunct{\mcitedefaultmidpunct}
{\mcitedefaultendpunct}{\mcitedefaultseppunct}\relax
\EndOfBibitem
\bibitem{ALICE:2021mgk}
ALICE collaboration, S.~Acharya {\em et~al.},
  \ifthenelse{\boolean{articletitles}}{\emph{{Measurement of beauty and charm
  production in pp collisions at $ \sqrt{s} $ = 5.02 TeV via non-prompt and
  prompt D mesons}}, }{}\href{https://doi.org/10.1007/JHEP05(2021)220}{JHEP
  \textbf{05} (2021) 220},
  \href{http://arxiv.org/abs/2102.13601}{{\normalfont\ttfamily
  arXiv:2102.13601}}\relax
\mciteBstWouldAddEndPuncttrue
\mciteSetBstMidEndSepPunct{\mcitedefaultmidpunct}
{\mcitedefaultendpunct}{\mcitedefaultseppunct}\relax
\EndOfBibitem
\bibitem{Andersson:1983ia}
B.~Andersson, G.~Gustafson, G.~Ingelman, and T.~Sjostrand,
  \ifthenelse{\boolean{articletitles}}{\emph{{Parton fragmentation and string
  dynamics}}, }{}\href{https://doi.org/10.1016/0370-1573(83)90080-7}{Phys.\
  Rept.\  \textbf{97} (1983) 31}\relax
\mciteBstWouldAddEndPuncttrue
\mciteSetBstMidEndSepPunct{\mcitedefaultmidpunct}
{\mcitedefaultendpunct}{\mcitedefaultseppunct}\relax
\EndOfBibitem
\bibitem{Webber:1983if}
B.~R. Webber, \ifthenelse{\boolean{articletitles}}{\emph{{A QCD model for jet
  fragmentation including soft gluon interference}},
  }{}\href{https://doi.org/10.1016/0550-3213(84)90333-X}{Nucl.\ Phys.\
  \textbf{B238} (1984) 492}\relax
\mciteBstWouldAddEndPuncttrue
\mciteSetBstMidEndSepPunct{\mcitedefaultmidpunct}
{\mcitedefaultendpunct}{\mcitedefaultseppunct}\relax
\EndOfBibitem
\bibitem{CLEO:2005pyn}
CLEO collaboration, M.~Artuso {\em et~al.},
  \ifthenelse{\boolean{articletitles}}{\emph{{First evidence and measurement of
  $B_{s}^{(*)} \overline{B_{s}}^{(*)}$ production at the $\Upsilon(5S)$}},
  }{}\href{https://doi.org/10.1103/PhysRevLett.95.261801}{Phys.\ Rev.\ Lett.\
  \textbf{95} (2005) 261801},
  \href{http://arxiv.org/abs/hep-ex/0508047}{{\normalfont\ttfamily
  arXiv:hep-ex/0508047}}\relax
\mciteBstWouldAddEndPuncttrue
\mciteSetBstMidEndSepPunct{\mcitedefaultmidpunct}
{\mcitedefaultendpunct}{\mcitedefaultseppunct}\relax
\EndOfBibitem
\bibitem{Belle:2006jvm}
Belle collaboration, A.~Drutskoy {\em et~al.},
  \ifthenelse{\boolean{articletitles}}{\emph{{Measurement of inclusive $D_{s}$,
  $D^{0}$ and $\jpsi$ rates and determination of the $B_{s}^{(*)}
  \overline{B_{s}}^{(*)}$ production fraction in $b\overline{b}$ events at the
  $\Upsilon(5S)$ resonance}},
  }{}\href{https://doi.org/10.1103/PhysRevLett.98.052001}{Phys.\ Rev.\ Lett.\
  \textbf{98} (2007) 052001},
  \href{http://arxiv.org/abs/hep-ex/0608015}{{\normalfont\ttfamily
  arXiv:hep-ex/0608015}}\relax
\mciteBstWouldAddEndPuncttrue
\mciteSetBstMidEndSepPunct{\mcitedefaultmidpunct}
{\mcitedefaultendpunct}{\mcitedefaultseppunct}\relax
\EndOfBibitem
\bibitem{CLEO:2006dkk}
CLEO collaboration, G.~S. Huang {\em et~al.},
  \ifthenelse{\boolean{articletitles}}{\emph{{Measurement of
  $\BR(\Upsilon(5S)\rightarrow B_{s}^{(*)} \overline{B_{s}}^{(*)})$ using
  $\phi$ mesons}}, }{}\href{https://doi.org/10.1103/PhysRevD.75.012002}{Phys.\
  Rev.\  \textbf{D75} (2007) 012002},
  \href{http://arxiv.org/abs/hep-ex/0610035}{{\normalfont\ttfamily
  arXiv:hep-ex/0610035}}\relax
\mciteBstWouldAddEndPuncttrue
\mciteSetBstMidEndSepPunct{\mcitedefaultmidpunct}
{\mcitedefaultendpunct}{\mcitedefaultseppunct}\relax
\EndOfBibitem
\bibitem{OPAL:1992zsd}
OPAL collaboration, P.~D. Acton {\em et~al.},
  \ifthenelse{\boolean{articletitles}}{\emph{{Evidence for the existence of the
  strange $b$ flavored meson $B_s^0$ in $Z^0$ decays}},
  }{}\href{https://doi.org/10.1016/0370-2693(92)91578-W}{Phys.\ Lett.\
  \textbf{B295} (1992) 357}\relax
\mciteBstWouldAddEndPuncttrue
\mciteSetBstMidEndSepPunct{\mcitedefaultmidpunct}
{\mcitedefaultendpunct}{\mcitedefaultseppunct}\relax
\EndOfBibitem
\bibitem{ALEPH:1995npv}
ALEPH collaboration, D.~Buskulic {\em et~al.},
  \ifthenelse{\boolean{articletitles}}{\emph{{Measurement of the $\Bs$ lifetime
  and production rate with $D_s - \ell^+$ combinations in $Z$ decays}},
  }{}\href{https://doi.org/10.1016/0370-2693(95)01173-N}{Phys.\ Lett.\
  \textbf{B361} (1995) 221}\relax
\mciteBstWouldAddEndPuncttrue
\mciteSetBstMidEndSepPunct{\mcitedefaultmidpunct}
{\mcitedefaultendpunct}{\mcitedefaultseppunct}\relax
\EndOfBibitem
\bibitem{L3:2000gcz}
L3 collaboration, M.~Acciarri {\em et~al.},
  \ifthenelse{\boolean{articletitles}}{\emph{{Measurements of the $b \bar{b}$
  production cross-section and forward backward asymmetry at center-of-mass
  energies above the $Z$ pole at LEP}},
  }{}\href{https://doi.org/10.1016/S0370-2693(00)00676-6}{Phys.\ Lett.\
  \textbf{B485} (2000) 71},
  \href{http://arxiv.org/abs/hep-ex/0005023}{{\normalfont\ttfamily
  arXiv:hep-ex/0005023}}\relax
\mciteBstWouldAddEndPuncttrue
\mciteSetBstMidEndSepPunct{\mcitedefaultmidpunct}
{\mcitedefaultendpunct}{\mcitedefaultseppunct}\relax
\EndOfBibitem
\bibitem{DELPHI:2003pao}
DELPHI collaboration, J.~Abdallah {\em et~al.},
  \ifthenelse{\boolean{articletitles}}{\emph{{A measurement of the branching
  fractions of the $b$ quark into charged and neutral $b$ hadrons}},
  }{}\href{https://doi.org/10.1016/j.physletb.2003.09.070}{Phys.\ Lett.\
  \textbf{B576} (2003) 29},
  \href{http://arxiv.org/abs/hep-ex/0311005}{{\normalfont\ttfamily
  arXiv:hep-ex/0311005}}\relax
\mciteBstWouldAddEndPuncttrue
\mciteSetBstMidEndSepPunct{\mcitedefaultmidpunct}
{\mcitedefaultendpunct}{\mcitedefaultseppunct}\relax
\EndOfBibitem
\bibitem{Collins:1989gx}
J.~C. Collins, D.~E. Soper, and G.~F. Sterman,
  \ifthenelse{\boolean{articletitles}}{\emph{{Factorization of hard processes
  in QCD}}, }{}\href{https://doi.org/10.1142/9789814503266_0001}{Adv.\ Ser.\
  Direct.\ High Energy Phys.\  \textbf{5} (1989) 1},
  \href{http://arxiv.org/abs/hep-ph/0409313}{{\normalfont\ttfamily
  arXiv:hep-ph/0409313}}\relax
\mciteBstWouldAddEndPuncttrue
\mciteSetBstMidEndSepPunct{\mcitedefaultmidpunct}
{\mcitedefaultendpunct}{\mcitedefaultseppunct}\relax
\EndOfBibitem
\bibitem{CDF:1996ivk}
CDF collaboration, F.~Abe {\em et~al.},
  \ifthenelse{\boolean{articletitles}}{\emph{{Ratios of bottom meson branching
  fractions involving $J/\psi$ mesons and determination of $b$ quark
  fragmentation fractions}},
  }{}\href{https://doi.org/10.1103/PhysRevD.54.6596}{Phys.\ Rev.\ D \textbf{54}
  (1996) 6596},
  \href{http://arxiv.org/abs/hep-ex/9607003}{{\normalfont\ttfamily
  arXiv:hep-ex/9607003}}\relax
\mciteBstWouldAddEndPuncttrue
\mciteSetBstMidEndSepPunct{\mcitedefaultmidpunct}
{\mcitedefaultendpunct}{\mcitedefaultseppunct}\relax
\EndOfBibitem
\bibitem{LHCb-PAPER-2012-037}
LHCb collaboration, R.~Aaij {\em et~al.},
  \ifthenelse{\boolean{articletitles}}{\emph{{Measurement of the fragmentation
  fraction ratio $f_s/f_d$ and its dependence on \B meson kinematics}},
  }{}\href{https://doi.org/10.1007/JHEP04(2013)001}{JHEP \textbf{04} (2013)
  001}, \href{http://arxiv.org/abs/1301.5286}{{\normalfont\ttfamily
  arXiv:1301.5286}}\relax
\mciteBstWouldAddEndPuncttrue
\mciteSetBstMidEndSepPunct{\mcitedefaultmidpunct}
{\mcitedefaultendpunct}{\mcitedefaultseppunct}\relax
\EndOfBibitem
\bibitem{ATLAS:2015esn}
ATLAS collaboration, G.~Aad {\em et~al.},
  \ifthenelse{\boolean{articletitles}}{\emph{{Determination of the ratio of
  $b$-quark fragmentation fractions $f_s/f_d$ in $pp$ collisions at
  $\sqrt{s}=7$ TeV with the ATLAS detector}},
  }{}\href{https://doi.org/10.1103/PhysRevLett.115.262001}{Phys.\ Rev.\ Lett.\
  \textbf{115} (2015) 262001},
  \href{http://arxiv.org/abs/1507.08925}{{\normalfont\ttfamily
  arXiv:1507.08925}}\relax
\mciteBstWouldAddEndPuncttrue
\mciteSetBstMidEndSepPunct{\mcitedefaultmidpunct}
{\mcitedefaultendpunct}{\mcitedefaultseppunct}\relax
\EndOfBibitem
\bibitem{LHCb-PAPER-2019-020}
LHCb collaboration, R.~Aaij {\em et~al.},
  \ifthenelse{\boolean{articletitles}}{\emph{{Measurement of $f_s / f_u$
  variation with proton-proton collision energy and \B-meson kinematics}},
  }{}\href{https://doi.org/10.1103/PhysRevLett.124.122002}{Phys.\ Rev.\ Lett.\
  \textbf{124} (2020) 122002},
  \href{http://arxiv.org/abs/1910.09934}{{\normalfont\ttfamily
  arXiv:1910.09934}}\relax
\mciteBstWouldAddEndPuncttrue
\mciteSetBstMidEndSepPunct{\mcitedefaultmidpunct}
{\mcitedefaultendpunct}{\mcitedefaultseppunct}\relax
\EndOfBibitem
\bibitem{LHCb-PAPER-2020-046}
LHCb collaboration, R.~Aaij {\em et~al.},
  \ifthenelse{\boolean{articletitles}}{\emph{{Precise measurement of the
  $f_s/f_d$ ratio of fragmentation fractions and of $B^0_s$ decay branching
  fractions}}, }{}\href{https://doi.org/10.1103/PhysRevD.104.032005}{Phys.\
  Rev.\  \textbf{D104} (2021) 032005},
  \href{http://arxiv.org/abs/2103.06810}{{\normalfont\ttfamily
  arXiv:2103.06810}}\relax
\mciteBstWouldAddEndPuncttrue
\mciteSetBstMidEndSepPunct{\mcitedefaultmidpunct}
{\mcitedefaultendpunct}{\mcitedefaultseppunct}\relax
\EndOfBibitem
\bibitem{LHCb-PAPER-2011-018}
LHCb collaboration, R.~Aaij {\em et~al.},
  \ifthenelse{\boolean{articletitles}}{\emph{{Measurement of \bquark hadron
  production fractions in 7\tev\ \proton\proton collisions}},
  }{}\href{https://doi.org/10.1103/PhysRevD.85.032008}{Phys.\ Rev.\
  \textbf{D85} (2012) 032008},
  \href{http://arxiv.org/abs/1111.2357}{{\normalfont\ttfamily
  arXiv:1111.2357}}\relax
\mciteBstWouldAddEndPuncttrue
\mciteSetBstMidEndSepPunct{\mcitedefaultmidpunct}
{\mcitedefaultendpunct}{\mcitedefaultseppunct}\relax
\EndOfBibitem
\bibitem{LHCb-PAPER-2018-050}
LHCb collaboration, R.~Aaij {\em et~al.},
  \ifthenelse{\boolean{articletitles}}{\emph{{Measurement of $b$-hadron
  fractions in 13\tev\ \proton\proton collisions}},
  }{}\href{https://doi.org/10.1103/PhysRevD.100.031102}{Phys.\ Rev.\
  \textbf{D100} (2019) 031102(R)},
  \href{http://arxiv.org/abs/1902.06794}{{\normalfont\ttfamily
  arXiv:1902.06794}}\relax
\mciteBstWouldAddEndPuncttrue
\mciteSetBstMidEndSepPunct{\mcitedefaultmidpunct}
{\mcitedefaultendpunct}{\mcitedefaultseppunct}\relax
\EndOfBibitem
\bibitem{ALICE:2021dhb}
ALICE collaboration, S.~Acharya {\em et~al.},
  \ifthenelse{\boolean{articletitles}}{\emph{{Charm-quark fragmentation
  fractions and production cross section at midrapidity in pp collisions at the
  LHC}}, }{}\href{https://doi.org/10.1103/PhysRevD.105.L011103}{Phys.\ Rev.\
  \textbf{D105} (2022) L011103},
  \href{http://arxiv.org/abs/2105.06335}{{\normalfont\ttfamily
  arXiv:2105.06335}}\relax
\mciteBstWouldAddEndPuncttrue
\mciteSetBstMidEndSepPunct{\mcitedefaultmidpunct}
{\mcitedefaultendpunct}{\mcitedefaultseppunct}\relax
\EndOfBibitem
\bibitem{Berezhnoy:2013qln}
A.~V. Berezhnoy and A.~K. Likhoded,
  \ifthenelse{\boolean{articletitles}}{\emph{{Relative yield of heavy hadrons
  as a function of the transverse momentum in LHC experiments}},
  }{}\href{https://doi.org/10.1134/S1063778815020106}{Phys.\ Atom.\ Nucl.\
  \textbf{78} (2015) 292},
  \href{http://arxiv.org/abs/1309.1979}{{\normalfont\ttfamily
  arXiv:1309.1979}}\relax
\mciteBstWouldAddEndPuncttrue
\mciteSetBstMidEndSepPunct{\mcitedefaultmidpunct}
{\mcitedefaultendpunct}{\mcitedefaultseppunct}\relax
\EndOfBibitem
\bibitem{Vogt:1994zf}
R.~Vogt and S.~J. Brodsky, \ifthenelse{\boolean{articletitles}}{\emph{{QCD and
  intrinsic heavy quark predictions for leading charm and beauty
  hadroproduction}},
  }{}\href{https://doi.org/10.1016/0550-3213(94)00543-N}{Nucl.\ Phys.\
  \textbf{B438} (1995) 261},
  \href{http://arxiv.org/abs/hep-ph/9405236}{{\normalfont\ttfamily
  arXiv:hep-ph/9405236}}\relax
\mciteBstWouldAddEndPuncttrue
\mciteSetBstMidEndSepPunct{\mcitedefaultmidpunct}
{\mcitedefaultendpunct}{\mcitedefaultseppunct}\relax
\EndOfBibitem
\bibitem{Braaten:2002yt}
E.~Braaten, Y.~Jia, and T.~Mehen,
  \ifthenelse{\boolean{articletitles}}{\emph{{The leading particle effect from
  heavy quark recombination}},
  }{}\href{https://doi.org/10.1103/PhysRevLett.89.122002}{Phys.\ Rev.\ Lett.\
  \textbf{89} (2002) 122002},
  \href{http://arxiv.org/abs/hep-ph/0205149}{{\normalfont\ttfamily
  arXiv:hep-ph/0205149}}\relax
\mciteBstWouldAddEndPuncttrue
\mciteSetBstMidEndSepPunct{\mcitedefaultmidpunct}
{\mcitedefaultendpunct}{\mcitedefaultseppunct}\relax
\EndOfBibitem
\bibitem{Rapp:2003wn}
R.~Rapp and E.~V. Shuryak, \ifthenelse{\boolean{articletitles}}{\emph{{D meson
  production from recombination in hadronic collisions}},
  }{}\href{https://doi.org/10.1103/PhysRevD.67.074036}{Phys.\ Rev.\
  \textbf{D67} (2003) 074036},
  \href{http://arxiv.org/abs/hep-ph/0301245}{{\normalfont\ttfamily
  arXiv:hep-ph/0301245}}\relax
\mciteBstWouldAddEndPuncttrue
\mciteSetBstMidEndSepPunct{\mcitedefaultmidpunct}
{\mcitedefaultendpunct}{\mcitedefaultseppunct}\relax
\EndOfBibitem
\bibitem{He:2019vgs}
M.~He and R.~Rapp, \ifthenelse{\boolean{articletitles}}{\emph{{Hadronization
  and charm-hadron ratios in heavy-ion collisions}},
  }{}\href{https://doi.org/10.1103/PhysRevLett.124.042301}{Phys.\ Rev.\ Lett.\
  \textbf{124} (2020) 042301},
  \href{http://arxiv.org/abs/1905.09216}{{\normalfont\ttfamily
  arXiv:1905.09216}}\relax
\mciteBstWouldAddEndPuncttrue
\mciteSetBstMidEndSepPunct{\mcitedefaultmidpunct}
{\mcitedefaultendpunct}{\mcitedefaultseppunct}\relax
\EndOfBibitem
\bibitem{Minissale:2020bif}
V.~Minissale, S.~Plumari, and V.~Greco,
  \ifthenelse{\boolean{articletitles}}{\emph{{Charm hadrons in pp collisions at
  LHC energy within a coalescence plus fragmentation approach}},
  }{}\href{https://doi.org/10.1016/j.physletb.2021.136622}{Phys.\ Lett.\
  \textbf{B821} (2021) 136622},
  \href{http://arxiv.org/abs/2012.12001}{{\normalfont\ttfamily
  arXiv:2012.12001}}\relax
\mciteBstWouldAddEndPuncttrue
\mciteSetBstMidEndSepPunct{\mcitedefaultmidpunct}
{\mcitedefaultendpunct}{\mcitedefaultseppunct}\relax
\EndOfBibitem
\bibitem{Fries:2003vb}
R.~J. Fries, B.~Muller, C.~Nonaka, and S.~A. Bass,
  \ifthenelse{\boolean{articletitles}}{\emph{{Hadronization in heavy ion
  collisions: recombination and fragmentation of partons}},
  }{}\href{https://doi.org/10.1103/PhysRevLett.90.202303}{Phys.\ Rev.\ Lett.\
  \textbf{90} (2003) 202303},
  \href{http://arxiv.org/abs/nucl-th/0301087}{{\normalfont\ttfamily
  arXiv:nucl-th/0301087}}\relax
\mciteBstWouldAddEndPuncttrue
\mciteSetBstMidEndSepPunct{\mcitedefaultmidpunct}
{\mcitedefaultendpunct}{\mcitedefaultseppunct}\relax
\EndOfBibitem
\bibitem{Greco:2003xt}
V.~Greco, C.~M. Ko, and P.~Levai,
  \ifthenelse{\boolean{articletitles}}{\emph{{Parton coalescence and
  anti-proton / pion anomaly at RHIC}},
  }{}\href{https://doi.org/10.1103/PhysRevLett.90.202302}{Phys.\ Rev.\ Lett.\
  \textbf{90} (2003) 202302},
  \href{http://arxiv.org/abs/nucl-th/0301093}{{\normalfont\ttfamily
  arXiv:nucl-th/0301093}}\relax
\mciteBstWouldAddEndPuncttrue
\mciteSetBstMidEndSepPunct{\mcitedefaultmidpunct}
{\mcitedefaultendpunct}{\mcitedefaultseppunct}\relax
\EndOfBibitem
\bibitem{Molnar:2003ff}
D.~Molnar and S.~A. Voloshin,
  \ifthenelse{\boolean{articletitles}}{\emph{{Elliptic flow at large transverse
  momenta from quark coalescence}},
  }{}\href{https://doi.org/10.1103/PhysRevLett.91.092301}{Phys.\ Rev.\ Lett.\
  \textbf{91} (2003) 092301},
  \href{http://arxiv.org/abs/nucl-th/0302014}{{\normalfont\ttfamily
  arXiv:nucl-th/0302014}}\relax
\mciteBstWouldAddEndPuncttrue
\mciteSetBstMidEndSepPunct{\mcitedefaultmidpunct}
{\mcitedefaultendpunct}{\mcitedefaultseppunct}\relax
\EndOfBibitem
\bibitem{CMS:2018eso}
CMS collaboration, A.~M. Sirunyan {\em et~al.},
  \ifthenelse{\boolean{articletitles}}{\emph{{Measurement of B$^0_\mathrm{s}$
  meson production in pp and PbPb collisions at $\sqrt{s_\mathrm{NN}} =$ 5.02
  TeV}}, }{}\href{https://doi.org/10.1016/j.physletb.2019.07.014}{Phys.\ Lett.\
  B \textbf{796} (2019) 168},
  \href{http://arxiv.org/abs/1810.03022}{{\normalfont\ttfamily
  arXiv:1810.03022}}\relax
\mciteBstWouldAddEndPuncttrue
\mciteSetBstMidEndSepPunct{\mcitedefaultmidpunct}
{\mcitedefaultendpunct}{\mcitedefaultseppunct}\relax
\EndOfBibitem
\bibitem{CMS:2021mzx}
CMS collaboration, A.~Tumasyan {\em et~al.},
  \ifthenelse{\boolean{articletitles}}{\emph{{Observation of B$^0_\mathrm{s}$
  mesons and measurement of the B$^0_\mathrm{s}$/B$^{+}$ yield ratio in PbPb
  collisions at $\sqrt{s_\mathrm{NN}} =$ 5.02 TeV}},
  }{}\href{https://doi.org/10.1016/j.physletb.2022.137062}{Phys.\ Lett.\ B
  \textbf{829} (2022) 137062},
  \href{http://arxiv.org/abs/2109.01908}{{\normalfont\ttfamily
  arXiv:2109.01908}}\relax
\mciteBstWouldAddEndPuncttrue
\mciteSetBstMidEndSepPunct{\mcitedefaultmidpunct}
{\mcitedefaultendpunct}{\mcitedefaultseppunct}\relax
\EndOfBibitem
\bibitem{CMS:2010ifv}
CMS collaboration, V.~Khachatryan {\em et~al.},
  \ifthenelse{\boolean{articletitles}}{\emph{{Observation of long-range
  near-side angular correlations in proton-proton collisions at the LHC}},
  }{}\href{https://doi.org/10.1007/JHEP09(2010)091}{JHEP \textbf{09} (2010)
  091}, \href{http://arxiv.org/abs/1009.4122}{{\normalfont\ttfamily
  arXiv:1009.4122}}\relax
\mciteBstWouldAddEndPuncttrue
\mciteSetBstMidEndSepPunct{\mcitedefaultmidpunct}
{\mcitedefaultendpunct}{\mcitedefaultseppunct}\relax
\EndOfBibitem
\bibitem{ATLAS:2015hzw}
ATLAS collaboration, G.~Aad {\em et~al.},
  \ifthenelse{\boolean{articletitles}}{\emph{{Observation of long-range
  elliptic azimuthal anisotropies in $\sqrt{s}=$13 and 2.76 TeV $pp$ collisions
  with the ATLAS detector}},
  }{}\href{https://doi.org/10.1103/PhysRevLett.116.172301}{Phys.\ Rev.\ Lett.\
  \textbf{116} (2016) 172301},
  \href{http://arxiv.org/abs/1509.04776}{{\normalfont\ttfamily
  arXiv:1509.04776}}\relax
\mciteBstWouldAddEndPuncttrue
\mciteSetBstMidEndSepPunct{\mcitedefaultmidpunct}
{\mcitedefaultendpunct}{\mcitedefaultseppunct}\relax
\EndOfBibitem
\bibitem{CMS:2016fnw}
CMS collaboration, V.~Khachatryan {\em et~al.},
  \ifthenelse{\boolean{articletitles}}{\emph{{Evidence for collectivity in pp
  collisions at the LHC}},
  }{}\href{https://doi.org/10.1016/j.physletb.2016.12.009}{Phys.\ Lett.\
  \textbf{B65} (2017) 193},
  \href{http://arxiv.org/abs/1606.06198}{{\normalfont\ttfamily
  arXiv:1606.06198}}\relax
\mciteBstWouldAddEndPuncttrue
\mciteSetBstMidEndSepPunct{\mcitedefaultmidpunct}
{\mcitedefaultendpunct}{\mcitedefaultseppunct}\relax
\EndOfBibitem
\bibitem{ALICE:2017jyt}
ALICE collaboration, J.~Adam {\em et~al.},
  \ifthenelse{\boolean{articletitles}}{\emph{{Enhanced production of
  multi-strange hadrons in high-multiplicity proton-proton collisions}},
  }{}\href{https://doi.org/10.1038/nphys4111}{Nature Phys.\  \textbf{13} (2017)
  535}, \href{http://arxiv.org/abs/1606.07424}{{\normalfont\ttfamily
  arXiv:1606.07424}}\relax
\mciteBstWouldAddEndPuncttrue
\mciteSetBstMidEndSepPunct{\mcitedefaultmidpunct}
{\mcitedefaultendpunct}{\mcitedefaultseppunct}\relax
\EndOfBibitem
\bibitem{Koch:1986ud}
P.~Koch, B.~Muller, and J.~Rafelski,
  \ifthenelse{\boolean{articletitles}}{\emph{{Strangeness in relativistic heavy
  ion collisions}},
  }{}\href{https://doi.org/10.1016/0370-1573(86)90096-7}{Phys.\ Rept.\
  \textbf{142} (1986) 167}\relax
\mciteBstWouldAddEndPuncttrue
\mciteSetBstMidEndSepPunct{\mcitedefaultmidpunct}
{\mcitedefaultendpunct}{\mcitedefaultseppunct}\relax
\EndOfBibitem
\bibitem{LHCb-TDR-005}
LHCb collaboration, \ifthenelse{\boolean{articletitles}}{\emph{{LHCb VELO
  (VErtex LOcator): Technical Design Report}}, }{}
  \href{http://cdsweb.cern.ch/search?p=CERN-LHCC-2001-011&f=reportnumber&action_search=Search&c=LHCb}
  {CERN-LHCC-2001-011}, 2001\relax
\mciteBstWouldAddEndPuncttrue
\mciteSetBstMidEndSepPunct{\mcitedefaultmidpunct}
{\mcitedefaultendpunct}{\mcitedefaultseppunct}\relax
\EndOfBibitem
\bibitem{LHCb-DP-2014-001}
R.~Aaij {\em et~al.}, \ifthenelse{\boolean{articletitles}}{\emph{{Performance
  of the LHCb Vertex Locator}},
  }{}\href{https://doi.org/10.1088/1748-0221/9/09/P09007}{JINST \textbf{9}
  (2014) P09007}, \href{http://arxiv.org/abs/1405.7808}{{\normalfont\ttfamily
  arXiv:1405.7808}}\relax
\mciteBstWouldAddEndPuncttrue
\mciteSetBstMidEndSepPunct{\mcitedefaultmidpunct}
{\mcitedefaultendpunct}{\mcitedefaultseppunct}\relax
\EndOfBibitem
\bibitem{LHCb-DP-2008-001}
LHCb collaboration, A.~A. Alves~Jr.\ {\em et~al.},
  \ifthenelse{\boolean{articletitles}}{\emph{{The \lhcb detector at the LHC}},
  }{}\href{https://doi.org/10.1088/1748-0221/3/08/S08005}{JINST \textbf{3}
  (2008) S08005}\relax
\mciteBstWouldAddEndPuncttrue
\mciteSetBstMidEndSepPunct{\mcitedefaultmidpunct}
{\mcitedefaultendpunct}{\mcitedefaultseppunct}\relax
\EndOfBibitem
\bibitem{LHCb-DP-2014-002}
LHCb collaboration, R.~Aaij {\em et~al.},
  \ifthenelse{\boolean{articletitles}}{\emph{{LHCb detector performance}},
  }{}\href{https://doi.org/10.1142/S0217751X15300227}{Int.\ J.\ Mod.\ Phys.\
  \textbf{A30} (2015) 1530022},
  \href{http://arxiv.org/abs/1412.6352}{{\normalfont\ttfamily
  arXiv:1412.6352}}\relax
\mciteBstWouldAddEndPuncttrue
\mciteSetBstMidEndSepPunct{\mcitedefaultmidpunct}
{\mcitedefaultendpunct}{\mcitedefaultseppunct}\relax
\EndOfBibitem
\bibitem{Hulsbergen:2005pu}
W.~D. Hulsbergen, \ifthenelse{\boolean{articletitles}}{\emph{{Decay chain
  fitting with a Kalman filter}},
  }{}\href{https://doi.org/10.1016/j.nima.2005.06.078}{Nucl.\ Instrum.\ Meth.\
  \textbf{A552} (2005) 566},
  \href{http://arxiv.org/abs/physics/0503191}{{\normalfont\ttfamily
  arXiv:physics/0503191}}\relax
\mciteBstWouldAddEndPuncttrue
\mciteSetBstMidEndSepPunct{\mcitedefaultmidpunct}
{\mcitedefaultendpunct}{\mcitedefaultseppunct}\relax
\EndOfBibitem
\bibitem{Sjostrand:2007gs}
T.~Sj\"{o}strand, S.~Mrenna, and P.~Skands,
  \ifthenelse{\boolean{articletitles}}{\emph{{A brief introduction to PYTHIA
  8.1}}, }{}\href{https://doi.org/10.1016/j.cpc.2008.01.036}{Comput.\ Phys.\
  Commun.\  \textbf{178} (2008) 852},
  \href{http://arxiv.org/abs/0710.3820}{{\normalfont\ttfamily
  arXiv:0710.3820}}\relax
\mciteBstWouldAddEndPuncttrue
\mciteSetBstMidEndSepPunct{\mcitedefaultmidpunct}
{\mcitedefaultendpunct}{\mcitedefaultseppunct}\relax
\EndOfBibitem
\bibitem{Sjostrand:2006za}
T.~Sj\"{o}strand, S.~Mrenna, and P.~Skands,
  \ifthenelse{\boolean{articletitles}}{\emph{{PYTHIA 6.4 physics and manual}},
  }{}\href{https://doi.org/10.1088/1126-6708/2006/05/026}{JHEP \textbf{05}
  (2006) 026}, \href{http://arxiv.org/abs/hep-ph/0603175}{{\normalfont\ttfamily
  arXiv:hep-ph/0603175}}\relax
\mciteBstWouldAddEndPuncttrue
\mciteSetBstMidEndSepPunct{\mcitedefaultmidpunct}
{\mcitedefaultendpunct}{\mcitedefaultseppunct}\relax
\EndOfBibitem
\bibitem{LHCb-PROC-2010-056}
I.~Belyaev {\em et~al.}, \ifthenelse{\boolean{articletitles}}{\emph{{Handling
  of the generation of primary events in Gauss, the LHCb simulation
  framework}}, }{}\href{https://doi.org/10.1088/1742-6596/331/3/032047}{J.\
  Phys.\ Conf.\ Ser.\  \textbf{331} (2011) 032047}\relax
\mciteBstWouldAddEndPuncttrue
\mciteSetBstMidEndSepPunct{\mcitedefaultmidpunct}
{\mcitedefaultendpunct}{\mcitedefaultseppunct}\relax
\EndOfBibitem
\bibitem{Lange:2001uf}
D.~J. Lange, \ifthenelse{\boolean{articletitles}}{\emph{{The EvtGen particle
  decay simulation package}},
  }{}\href{https://doi.org/10.1016/S0168-9002(01)00089-4}{Nucl.\ Instrum.\
  Meth.\  \textbf{A462} (2001) 152}\relax
\mciteBstWouldAddEndPuncttrue
\mciteSetBstMidEndSepPunct{\mcitedefaultmidpunct}
{\mcitedefaultendpunct}{\mcitedefaultseppunct}\relax
\EndOfBibitem
\bibitem{Allison:2006ve}
Geant4 collaboration, J.~Allison {\em et~al.},
  \ifthenelse{\boolean{articletitles}}{\emph{{Geant4 developments and
  applications}}, }{}\href{https://doi.org/10.1109/TNS.2006.869826}{IEEE
  Trans.\ Nucl.\ Sci.\  \textbf{53} (2006) 270}\relax
\mciteBstWouldAddEndPuncttrue
\mciteSetBstMidEndSepPunct{\mcitedefaultmidpunct}
{\mcitedefaultendpunct}{\mcitedefaultseppunct}\relax
\EndOfBibitem
\bibitem{Agostinelli:2002hh}
Geant4 collaboration, S.~Agostinelli {\em et~al.},
  \ifthenelse{\boolean{articletitles}}{\emph{{Geant4: A simulation toolkit}},
  }{}\href{https://doi.org/10.1016/S0168-9002(03)01368-8}{Nucl.\ Instrum.\
  Meth.\  \textbf{A506} (2003) 250}\relax
\mciteBstWouldAddEndPuncttrue
\mciteSetBstMidEndSepPunct{\mcitedefaultmidpunct}
{\mcitedefaultendpunct}{\mcitedefaultseppunct}\relax
\EndOfBibitem
\bibitem{LHCb-PROC-2011-006}
M.~Clemencic {\em et~al.}, \ifthenelse{\boolean{articletitles}}{\emph{{The
  \lhcb simulation application, Gauss: Design, evolution and experience}},
  }{}\href{https://doi.org/10.1088/1742-6596/331/3/032023}{J.\ Phys.\ Conf.\
  Ser.\  \textbf{331} (2011) 032023}\relax
\mciteBstWouldAddEndPuncttrue
\mciteSetBstMidEndSepPunct{\mcitedefaultmidpunct}
{\mcitedefaultendpunct}{\mcitedefaultseppunct}\relax
\EndOfBibitem
\bibitem{splot}
M.~Pivk and F.~R. Le~Diberder,
  \ifthenelse{\boolean{articletitles}}{\emph{{SPlot: a statistical tool to
  unfold data distributions}},
  }{}\href{https://doi.org/10.1016/j.nima.2005.08.106}{Nucl.\ Instrum.\ Meth.\
  \textbf{\bf{A}555} (2005) 356},
  \href{http://arxiv.org/abs/physics/0402083}{{\normalfont\ttfamily
  arXiv:physics/0402083}}\relax
\mciteBstWouldAddEndPuncttrue
\mciteSetBstMidEndSepPunct{\mcitedefaultmidpunct}
{\mcitedefaultendpunct}{\mcitedefaultseppunct}\relax
\EndOfBibitem
\bibitem{ParticleDataGroup:2020ssz}
Particle Data Group, P.~A. Zyla {\em et~al.},
  \ifthenelse{\boolean{articletitles}}{\emph{{Review of Particle Physics}},
  }{}\href{https://doi.org/10.1093/ptep/ptaa104}{PTEP \textbf{2020} (2020)
  083C01}\relax
\mciteBstWouldAddEndPuncttrue
\mciteSetBstMidEndSepPunct{\mcitedefaultmidpunct}
{\mcitedefaultendpunct}{\mcitedefaultseppunct}\relax
\EndOfBibitem
\bibitem{LHCb-PUB-2014-039}
S.~Tolk, J.~Albrecht, F.~Dettori, and A.~Pellegrino,
  \ifthenelse{\boolean{articletitles}}{\emph{{Data driven trigger efficiency
  determination at LHCb}}, }{}
  \href{http://cdsweb.cern.ch/search?p=LHCb-PUB-2014-039&f=reportnumber&action_search=Search&c=LHCb+Notes}
  {LHCb-PUB-2014-039}, 2014\relax
\mciteBstWouldAddEndPuncttrue
\mciteSetBstMidEndSepPunct{\mcitedefaultmidpunct}
{\mcitedefaultendpunct}{\mcitedefaultseppunct}\relax
\EndOfBibitem
\bibitem{LHCb-PAPER-2013-069}
LHCb collaboration, R.~Aaij {\em et~al.},
  \ifthenelse{\boolean{articletitles}}{\emph{{Measurement of resonant and \CP
  components in \mbox{\decay{\Bsb}{\jpsi\pip\pim}} decays}},
  }{}\href{https://doi.org/10.1103/PhysRevD.89.092006}{Phys.\ Rev.\
  \textbf{D89} (2014) 092006},
  \href{http://arxiv.org/abs/1402.6248}{{\normalfont\ttfamily
  arXiv:1402.6248}}\relax
\mciteBstWouldAddEndPuncttrue
\mciteSetBstMidEndSepPunct{\mcitedefaultmidpunct}
{\mcitedefaultendpunct}{\mcitedefaultseppunct}\relax
\EndOfBibitem
\bibitem{LHCb-PAPER-2014-012}
LHCb collaboration, R.~Aaij {\em et~al.},
  \ifthenelse{\boolean{articletitles}}{\emph{{Measurement of the resonant and
  \CP components in \mbox{\decay{\Bzb}{\jpsi\pip\pim}} decays}},
  }{}\href{https://doi.org/10.1103/PhysRevD.90.012003}{Phys.\ Rev.\
  \textbf{D90} (2014) 012003},
  \href{http://arxiv.org/abs/1404.5673}{{\normalfont\ttfamily
  arXiv:1404.5673}}\relax
\mciteBstWouldAddEndPuncttrue
\mciteSetBstMidEndSepPunct{\mcitedefaultmidpunct}
{\mcitedefaultendpunct}{\mcitedefaultseppunct}\relax
\EndOfBibitem
\bibitem{ref:SuppMat}
\ifthenelse{\boolean{articletitles}}{\emph{{See Supplemental Material at [link
  inserted by publisher] for numerical values of cross section ratios}},
  }{}\relax
\mciteBstWouldAddEndPuncttrue
\mciteSetBstMidEndSepPunct{\mcitedefaultmidpunct}
{\mcitedefaultendpunct}{\mcitedefaultseppunct}\relax
\EndOfBibitem
\bibitem{HFLAV:2019otj}
HFLAV collaboration, Y.~S. Amhis {\em et~al.},
  \ifthenelse{\boolean{articletitles}}{\emph{{Averages of b-hadron, c-hadron,
  and $\tau $-lepton properties as of 2018}},
  }{}\href{https://doi.org/10.1140/epjc/s10052-020-8156-7}{Eur.\ Phys.\ J.\
  \textbf{C81} (2021) 226},
  \href{http://arxiv.org/abs/1909.12524}{{\normalfont\ttfamily
  arXiv:1909.12524}}\relax
\mciteBstWouldAddEndPuncttrue
\mciteSetBstMidEndSepPunct{\mcitedefaultmidpunct}
{\mcitedefaultendpunct}{\mcitedefaultseppunct}\relax
\EndOfBibitem
\bibitem{Sjostrand:1987su}
T.~Sjostrand and M.~van Zijl, \ifthenelse{\boolean{articletitles}}{\emph{{A
  Multiple Interaction Model for the Event Structure in Hadron Collisions}},
  }{}\href{https://doi.org/10.1103/PhysRevD.36.2019}{Phys.\ Rev.\ D \textbf{36}
  (1987) 2019}\relax
\mciteBstWouldAddEndPuncttrue
\mciteSetBstMidEndSepPunct{\mcitedefaultmidpunct}
{\mcitedefaultendpunct}{\mcitedefaultseppunct}\relax
\EndOfBibitem
\bibitem{Sjostrand:2013cya}
T.~Sj\"ostrand, \ifthenelse{\boolean{articletitles}}{\emph{{Colour reconnection
  and its effects on precise measurements at the LHC}}, }{} 2013,
  \href{http://arxiv.org/abs/1310.8073}{{\normalfont\ttfamily
  arXiv:1310.8073}}\relax
\mciteBstWouldAddEndPuncttrue
\mciteSetBstMidEndSepPunct{\mcitedefaultmidpunct}
{\mcitedefaultendpunct}{\mcitedefaultseppunct}\relax
\EndOfBibitem
\end{mcitethebibliography}
 
\newpage
\centerline
{\large\bf LHCb collaboration}
\begin
{flushleft}
\small
R.~Aaij$^{32}$\lhcborcid{0000-0003-0533-1952},
A.S.W.~Abdelmotteleb$^{50}$\lhcborcid{0000-0001-7905-0542},
C.~Abellan~Beteta$^{44}$,
F.~Abudin{\'e}n$^{50}$\lhcborcid{0000-0002-6737-3528},
T.~Ackernley$^{54}$\lhcborcid{0000-0002-5951-3498},
B.~Adeva$^{40}$\lhcborcid{0000-0001-9756-3712},
M.~Adinolfi$^{48}$\lhcborcid{0000-0002-1326-1264},
H.~Afsharnia$^{9}$,
C.~Agapopoulou$^{13}$\lhcborcid{0000-0002-2368-0147},
C.A.~Aidala$^{76}$\lhcborcid{0000-0001-9540-4988}
% ...

{\footnotesize \it

% ...
$^{9}$Universit{\'e} Clermont Auvergne, CNRS/IN2P3, LPC, Clermont-Ferrand, France\\
% ...
$^{13}$LPNHE, Sorbonne Universit{\'e}, Paris Diderot Sorbonne Paris Cit{\'e}, CNRS/IN2P3, Paris, France\\
% ...
$^{32}$Nikhef National Institute for Subatomic Physics, Amsterdam, Netherlands\\
% ...
$^{40}$Instituto Galego de F{\'\i}sica de Altas Enerx{\'\i}as (IGFAE), Universidade de Santiago de Compostela, Santiago de Compostela, Spain\\
% ...
$^{44}$Physik-Institut, Universit{\"a}t Z{\"u}rich, Z{\"u}rich, Switzerland\\
% ...
$^{48}$H.H. Wills Physics Laboratory, University of Bristol, Bristol, United Kingdom\\
% ...
$^{50}$Department of Physics, University of Warwick, Coventry, United Kingdom\\
% ...
$^{54}$Oliver Lodge Laboratory, University of Liverpool, Liverpool, United Kingdom\\
% ...
$^{62}$Syracuse University, Syracuse, NY, United States\\
% ...
$^{76}$University of Michigan, Ann Arbor, MI, United States, associated to $^{62}$\\
\bigskip
% ...
}
\end{flushleft}

\end{document}